\documentclass[aps,prl,twocolumn,showpacs,superscriptaddress]{revtex4-1}
\usepackage{graphicx}
\usepackage{amsmath}
\usepackage{textcomp}
\usepackage{color}
\usepackage{xcolor}
\usepackage{upgreek}
\usepackage{tcolorbox}
\begin{document}

\title{Chiroptical response of a single plasmonic nanohelix}
\author{Pawe{\l} Wo{\'z}niak}
\email{pawel.wozniak@mpl.mpg.de}
\affiliation{Max Planck Institute for the Science of Light, Erlangen, Germany}
\affiliation{Institute of Optics, Information and Photonics, Friedrich-Alexander-University Erlangen-Nuremberg, Erlangen, Germany}
\affiliation{University of Ottawa Centre for Extreme and Quantum Photonics, University of Ottawa, Ottawa, Canada}
\author{Israel De Leon}
\email{ideleon@itesm.mx}
\affiliation{School of Engineering and Sciences, Tecnol{\'o}gico de Monterrey, Monterrey, Mexico}
\affiliation{University of Ottawa Centre for Extreme and Quantum Photonics, University of Ottawa, Ottawa, Canada}
\author{Katja H{\"o}flich}
\affiliation{Nanoscale Structures and Microscopic Analysis, Helmholtz-Zentrum Berlin f{\"u}r Materialien und Energie, Berlin, Germany}
\author{Caspar~Haverkamp}
\affiliation{Nanoscale Structures and Microscopic Analysis, Helmholtz-Zentrum Berlin f{\"u}r Materialien und Energie, Berlin, Germany}
\author{Silke~Christiansen}
\affiliation{Nanoscale Structures and Microscopic Analysis, Helmholtz-Zentrum Berlin f{\"u}r Materialien und Energie, Berlin, Germany}
\author{Gerd Leuchs}
\affiliation{Max Planck Institute for the Science of Light, Erlangen, Germany}
\affiliation{Institute of Optics, Information and Photonics, Friedrich-Alexander-University Erlangen-Nuremberg, Erlangen, Germany}
\affiliation{University of Ottawa Centre for Extreme and Quantum Photonics, University of Ottawa, Ottawa, Canada}
\author{Peter Banzer}
\email{peter.banzer@mpl.mpg.de}
\affiliation{Max Planck Institute for the Science of Light, Erlangen, Germany}
\affiliation{Institute of Optics, Information and Photonics, Friedrich-Alexander-University Erlangen-Nuremberg, Erlangen, Germany}
\affiliation{University of Ottawa Centre for Extreme and Quantum Photonics, University of Ottawa, Ottawa, Canada}
\date{\today}

\begin{abstract}
{We investigate the chiroptical response of a single plasmonic nanohelix interacting with a weakly-focused circularly-polarized Gaussian beam. The optical scattering at the fundamental resonance is characterized experimentally, and the chiral behavior of the nanohelix is explained based on a multipole analysis. The angularly resolved emission of the excited nanohelix is verified experimentally and it validates the theoretical results. Further, we study the first higher-order resonance and explain the formation of chiral dipoles in both cases.}
\end{abstract}
\maketitle
%
\section{Introduction} 
\vspace{-0.5cm}
Chirality in nature is a property of certain molecules which allows them to interact differently with right- and left-handed circularly polarized light. Chiral molecules lack a plane of mirror symmetry in their molecular structure. They exist in two different forms (enantiomers), which are mirror images of each other. Intuitive examples of chiral molecules are those exhibiting a helical shape in which the sense of twist defines their handedness. Interaction of light with chiral molecules leads to interesting optical phenomena such as optical rotation and circular dichroism~\cite{barron_2009}, and have importance in, e.g., the life sciences and chemical analysis \cite{ranjbar_2009, hendry_2010,zhao_2014,jack_2016,zhao_2017}. Chiroptical effects of natural compounds are rather weak and typically occur in the ultraviolet which limits their applications. However, the advances in nanofabrication enabled the manufacture of artificial media of enhanced chiral response over the optical spectral range~\cite{wang_2009,valev_2013,oh_2015,wang_2016}. Such media, typically called chiral metamaterials, consist of chiral metamolecules, which take the form of sub-wavelength chiral particles or particle-clusters. They are engineered to exhibit a strong chiral response by controlling the geometry of the particles~\cite{kuwata_gonokami_2005,rogacheva_2006,gansel_2009} or clusters ~\cite{roman_velazquez_2003,fan_2010,fan_2011,kuzyk_2012,hentschel_2013}, their spatial arrangement~\cite{decker_2009,huttunen_2011,zhao_2012} or their material composition~\cite{banzer_2016}. Chiral optical phenomena can also occur in anisotropic achiral particles under oblique illumination such that the particle and the wave vector of light form a chiral triad, a phenomenon known as pseudo or extrinsic chirality~\cite{plum_2009,plum_2009_2,sersic_2011,de_leon_2015}. A variety of chiral metameterias have been proposed and demonstrated to enable strong chiroptical effects~\cite{kuwata_gonokami_2005,rogacheva_2006, gansel_2009,zhao_2012,cui_2014,fedotov_2006,pfeiffer_2014} and unusual phenomena such as negative refraction~\cite{pendry_2004,zhang_2009,plum_2009_3,wu_2010} and the repulsive Casimir force~\cite{zhao_2009}. Tunable and strongly chiral metamolecules are also desired for the realization of high-efficiency circular-polarization optical elements~\cite{gansel_2009,zhao_2012} and nanodevices for dynamic light manipulation~\cite{ren_2012,zhang_2012,pfeiffer_2014,de_leon_2015}. \\
As it relates to the lack of mirror symmetry of molecules (or metamolecules), chirality is usually associated with a three-dimensional geometry~\cite{barron_2009}. Among the various three-dimensional chiral structures which have been investigated, helix-shaped scatterers have received significant attention due to their strong, broadband and tunable chiroptical properties~\cite{passaseo_2017,esposito_2015,kosters_2017,gansel_2012,gibbs_2013,yang_2010,esposito_2015_2}. In particular, metallic nanohelices supporting localized surface plasmons at optical frequencies have been used to create metamaterials~\cite{kosters_2017,gibbs_2013,yang_2010,esposito_2015} and metafluids~\cite{kuzyk_2012,mark_2013,song_2013} of enhanced chirality. Despite the intense research on these structures, only a few reports have focused on the fundamental chiroptical properties of a single plasmonic nanohelix~\cite{frank_2013,larsen_2014,hu_2016}.\\
\begin{figure*}[t]
\centering 
\includegraphics[width=1\textwidth]{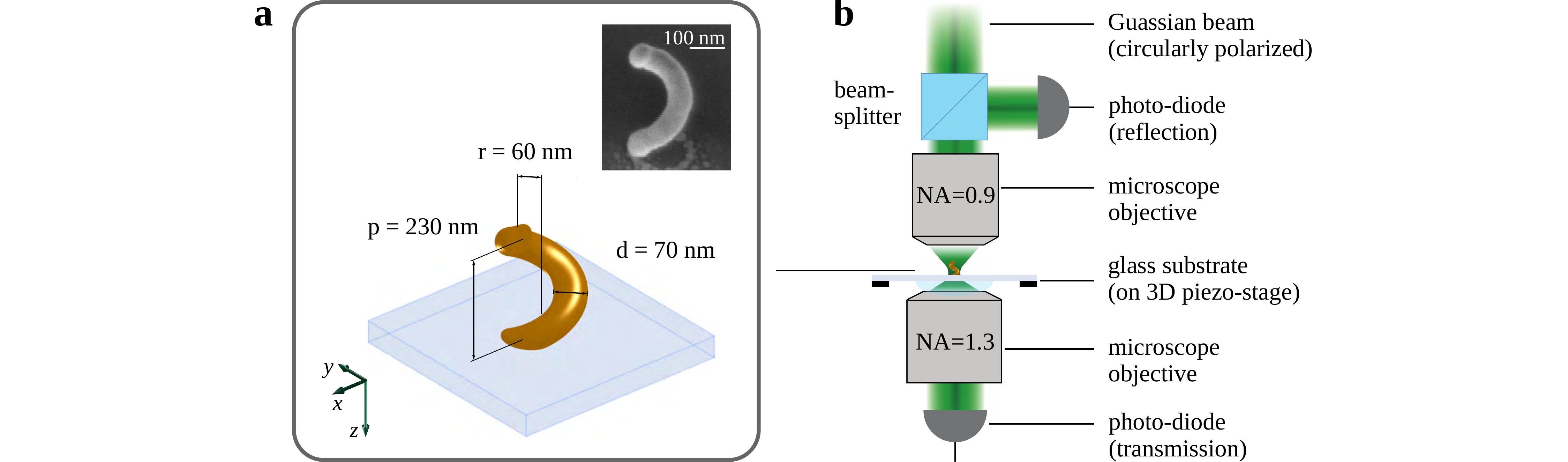} 
\caption{\textbf{Investigated nanostructure and experimental set-up.} (\textbf{a}) Schematic illustration and scanning-electron micrograph (inset) of the investigated gold-coated nanohelix on a glass substrate. (\textbf{b}) Simplified sketch of the experimental set-up utilized for the measurement of the nanohelix. A circularly polarized Gaussian beam is focused onto the nanohelix by a microscope objective with NA of 0.9. The incoming beam only partially fills the back focal plane of the focusing lens and decreases the effective focusing NA down to $0.5$. The same objective collects the reflected and the back-scattered light within a large solid angle equivalent to the NA of 0.9. Implementation of a 3D-piezo-stage allows for precise positioning of the structure on the optical axis in the focal plane. The transmitted and forward-scattered light is collected by a second immersion-type microscope objective with an NA of 1.3 which is located below the sample. In both directions, the total power is measured with photo-diodes.}
\label{fig:_schematics}
\end{figure*}
In this paper, we study experimentally and numerically the chiroptical response of a single-loop plasmonic nanohelix excited with circularly polarized light. The investigated structure is much smaller than the wavelength, both in the transverse and longitudinal directions. To this end, the nanohelix is characterized in terms of its optical scatting and its chiral signature is explained based on a multipole decomposition. The extracted spectra of the electric and magnetic dipole moments allow further to determine their contributions to the far-field scattering. Our results show that the optical response of the fundamental resonance has contributions from dipole and quadrupole moments, but it is dominated by the dipolar response.
\section{Chiroptical response of a single-loop nanohelix}
We investigate a right-handed nanohelix fabricated on a glass substrate using electron-beam-induced deposition and coated with gold (see Methods). The nanohelix is depicted schematically in Fig.~\ref{fig:_schematics}a. The structure exhibits nanoscopic dimensions with the radius of 60 nm, the pitch of 230 nm and the wire diameter of 70 nm. We characterize the chiroptical response of the nanohelix by measuring its reflectance and transmittance spectra, as well as its far-field radiation using a custom-built confocal-microscope setup (see simplified sketch Fig.~\ref{fig:_schematics}b). The helix is excited with a weakly focused circularly polarized Gaussian beam (effective focusing numerical aperture (NA) of 0.5). For that purpose, the incoming spectrally filtered beam partially fills the back focal plane of a high-NA lens (NA = 0.9). Note that this choice of the focusing and collection NA's provides a plan-wave-like illumination of the structure and yet collects the back-scattered and reflected light within a large solid angle. The transmitted and the forward-scattered light is collected by a second high-NA (1.3) immersion objective.  For the measurements, the nanohelix is positioned on the optical axis in the focal plane (see Methods and Ref.~\cite{banzer_2010,wozniak_2015} for more details).\\
The reflectance ($R$) and transmittance ($T$) spectra for right-handed circularly polarized (RCP) and left-handed circularly polarized (LCP) Gaussian excitation are measured wavelength-by-wavelength in the spectral range from 1300 nm to 1650 nm around the fundamental resonance of the investigated nanohelix. From this data, the absorbance $A$ can be calculated as $A=1-R-T$. Subsequently, the differential absorption, or equivalently, the circular dichroism $CD$ can be estimated as:
\begin{figure*}[t]
\centering 
\includegraphics[width=1\textwidth]{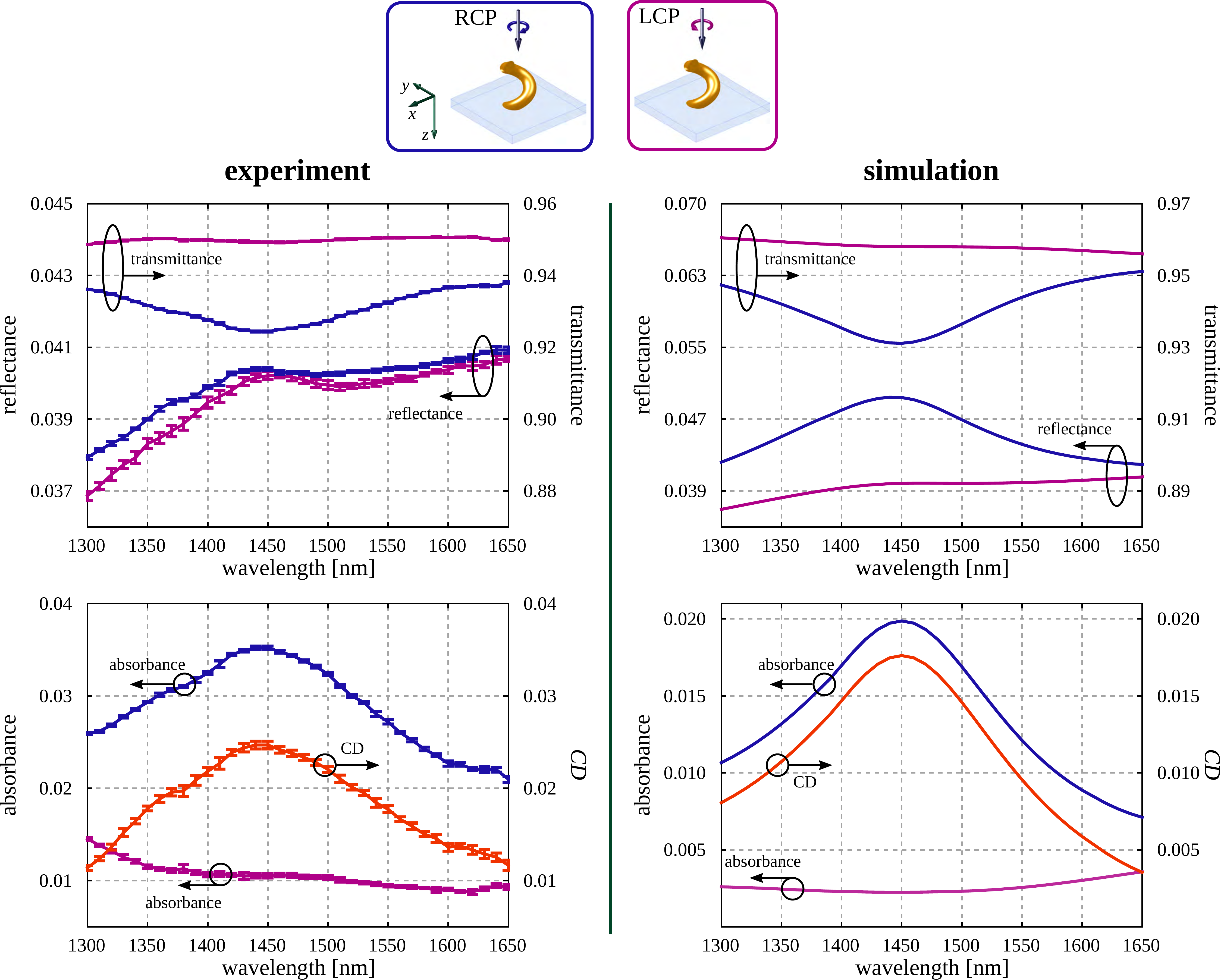} 
\caption{\textbf{Measurements and numerical simulations of the chiroptical response of a single plasmonic nanohelix.} Reflectance and transmittance spectra are shown at the top. The corresponding absorbance spectra for both polarization states of the incoming light are shown at the bottom. The resulting differential absorption ($CD$) is presented in the same graph.} 
\label{fig:_spectra}
\end{figure*}
\begin{equation*}
CD = A_\text{RCP}-A_\text{LCP}\text{.} 
\label{eq:_cd}
\end{equation*}
The corresponding datasets for $R$, $T$, $A$ and $CD$ are shown in Fig.~\ref{fig:_spectra}. Both the experimental and numerical data (based on the finite-difference time-domain (FDTD) method) show a resonance around 1450 nm for excitation with RCP light, matching the handedness of the nanohelix. For the incoming light of opposite handedness, however, there is no obvious fingerprint of resonant excitation. Thus, the resulting $CD$ spectrum presented in Fig.~\ref{fig:_spectra} (bottom) exhibits a broad peak centered at the wavelength of 1450 nm which is a clear sign of the optical activity of the nanohelix. The experimental and numerical spectra are in a good agreement with each other. Moreover, the FDTD simulations over a broader spectral range confirm that the observed $CD$ peak corresponds to the fundamental resonance of the structure (see Fig. S1). Hence, at $\lambda=1450$ nm the helix pitch (the largest dimension of the structure) is over six times smaller than the wavelength. The experimentally retrieved $CD$ reaches a maximum value of approximately 2.5 $\%$. It should be emphasized again that this value is recorded for a single nanohelix, in contrast to most of the studies reporting on $CD$ values for large arrays of chiral nanostructures \cite{gansel_2009,gansel_2010,gansel_2012,frank_2013,hu_2016,larsen_2014,esposito_2015_2}. In the case discussed here, a single nanohelix with an approximate diameter of only 120 nm is excited by a Gaussian beam focused weakly down to a spot with diameter of around 1.2 $\upmu$m (full width at half maximum at $\lambda = 1450$ nm).
\begin{figure*}[t]
\centering 
\includegraphics[width=1\textwidth]{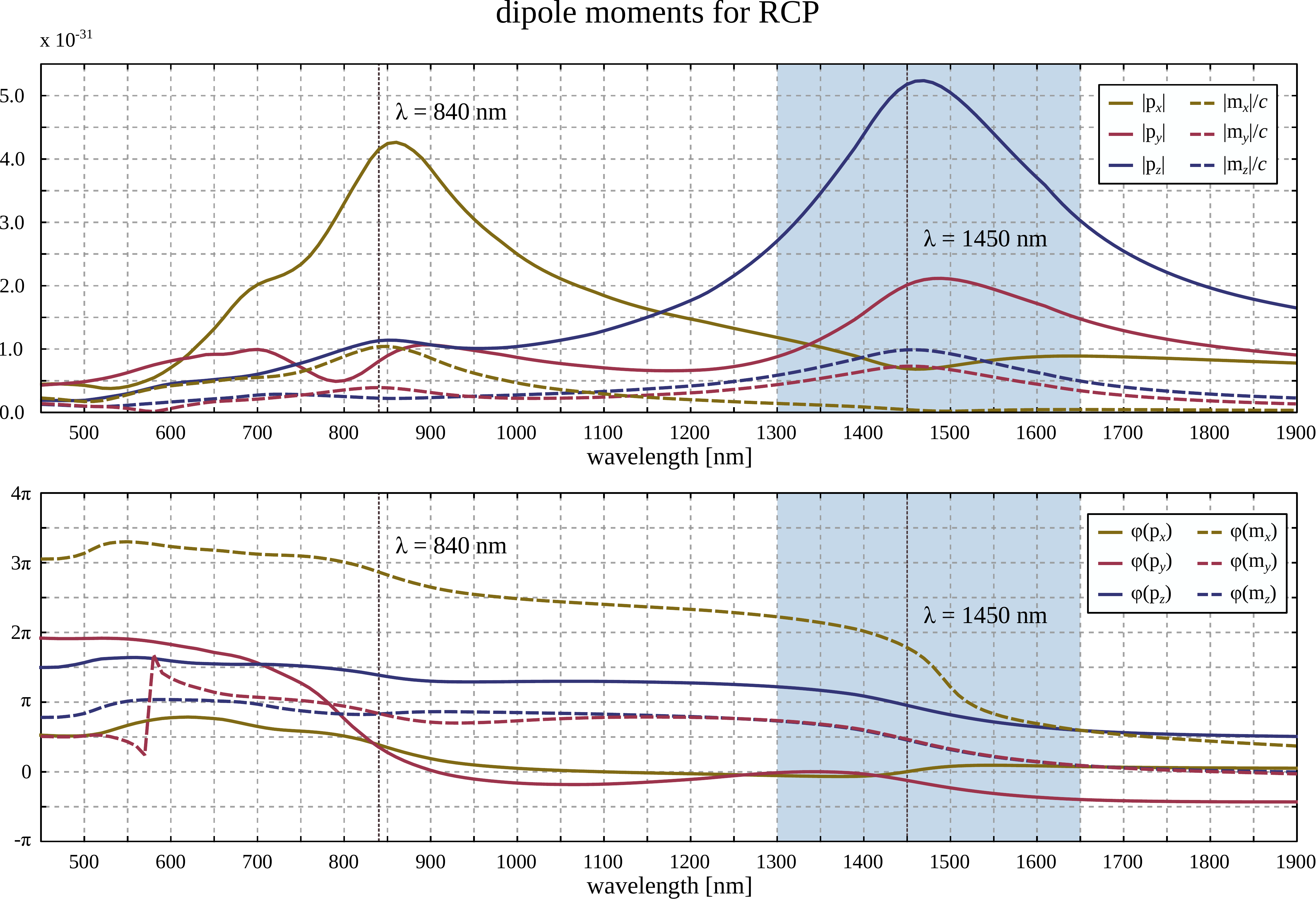}  
\caption{\textbf{FDTD-retrieved dipolar response of the nanohelix.} Absolute values and phases of the components of the electric $\textbf{p}$ and magnetic $\textbf{m}$ dipole moments for the right-handed helix under RCP illumination. The blue-shaded range spans over the spectral range of the experiment.}
\label{fig:_dipoles_rcp}
\end{figure*}
\section{Multipolar decomposition of the fundamental resonance}
\begin{figure*}[t]
\centering 
\includegraphics[width=1\textwidth]{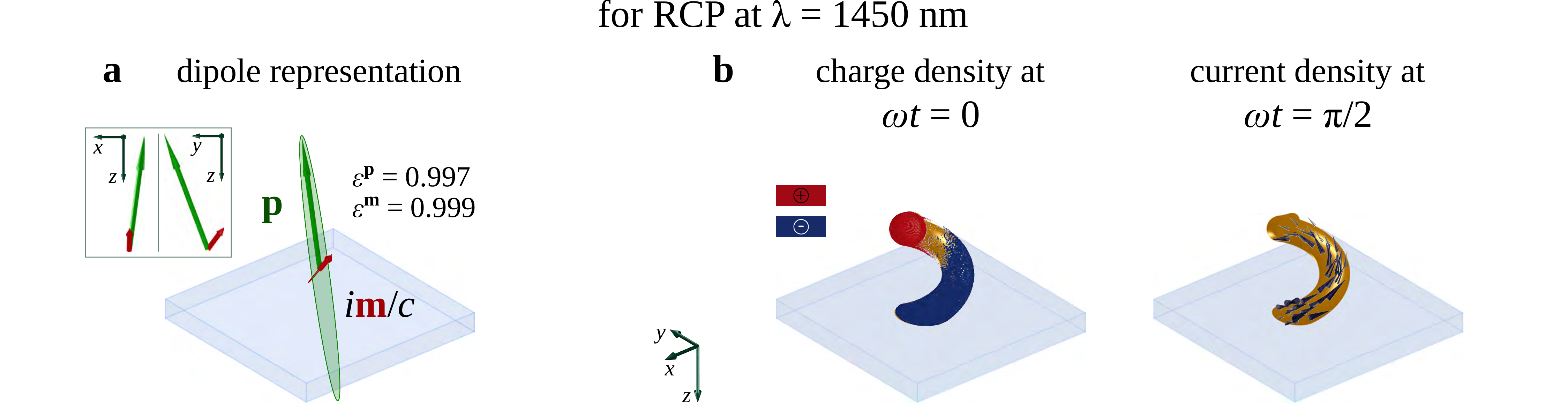}  
\caption{\textbf{Optical response of a single nanohelix at the wavelength of its fundamental resonance excited with right-handed circularly polarized light.} (\textbf{a}) The helix can be described as a system of coupled electric and magnetic dipoles of ellipticities $\epsilon$ and oscillating with a phase delay of $\pi/2$. Both dipoles have a strong component along the $z$-axis (the inset shows the projections of the retrieved dipoles on the $xz$- and $yz$-planes). (\textbf{b}) Distributions of the charge and current densities on the helix surface for two different snapshots in time. Effectively, the structure acts as a nanoLC-circuit accessible for only one handedness of the incoming light (here RCP) depending on its own geometrical twist.}
\label{fig:_resonance_1_rcp}
\end{figure*}

\begin{figure*}[t]
\centering 
\includegraphics[width=1\textwidth]{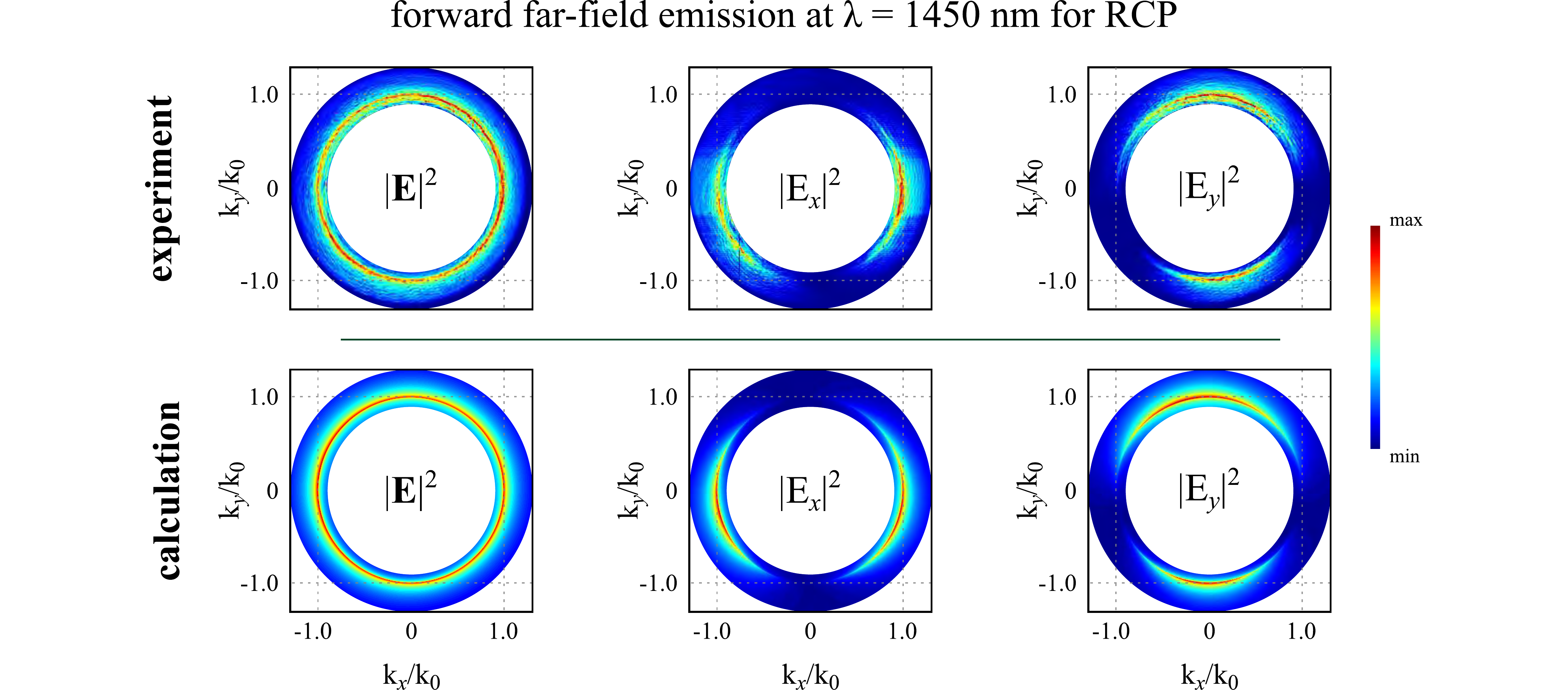} 
\caption{\textbf{Far-field emission of a single nanohelix at the wavelength of its fundamental resonance.} The forward-scattered light observed experimentally in the back focal plane is mostly radially polarized. The analytical prediction based on point-like dipole moments retrieved from the FDTD simulations near a dielectric substrate resembles well the experimentally acquired intensity patterns.}
\label{fig:_scattering}
\end{figure*}

To better understand the chiroptical response of the nanohelix, we extract the polarization density provided by the FDTD simulations and expand it into multipoles up to order 3 (dipoles, quadrupoles, and octupoles; see Methods). Because the nanohelix interacts effectively only with RCP, only this interaction is further investigated. Figure~\ref{fig:_dipoles_rcp} presents the spectra of the electric ($\textbf{p}$) and magnetic ($\textbf{m}$) dipole moments obtained with RCP excitation. The FDTD calculations were extended over the wavelengths between 450 nm and 1900 nm and show that, in the entire simulated spectral range, the optical response of the nanohelix is dominated by the two dipoles. Our multipolar analysis indicates that the fundamental resonance of the nanohelix has contributions of electric and magnetic dipoles and quadrupoles (see Supplementary Information). However, the contribution of the quadrupoles and any other higher order multipole can be neglected because their strength is several orders of magnitude smaller than the dipoles (a similar dipole-dominated chiroptical response of a two-loop helix was recently reported by Fruhnert \textit{et. al} in Ref.~\cite{fruhnert_2017}). Accordingly, at the resonance wavelength of 1450 nm, the two dipoles are elliptically polarized and oscillate predominantly along the helix axis ($z$-axis, see Fig.~\ref{fig:_resonance_1_rcp}a). The ellipticity $\epsilon$ was estimated to be 0.9965 and 0.9995 for the electric and magnetic dipoles, respectively. \\
For a more intuitive understanding of the dipolar composition, it is instructive to analyze the resonant charge and current distributions (see Fig.~\ref{fig:_resonance_1_rcp}b). At a certain point in time, the charges are well separated and accumulated at both ends of the helix. Such a dipolar distribution can be described as an electric dipole $\textbf{p}$, pointing from the center of the charge accumulation at one end of the helix to the other one. Hence, \textbf{p} is aligned predominantly along the helix axis having a small tilt from the $z$-axis which results from the finite length of the structure. The time-harmonic excitation, however, requires that after a certain time the charges migrate to the other side, giving rise to a current. Since the resonantly excited charges flow along a curved path, they induce a resonant magnetic field perpendicular to the current, resulting in a magnetic dipole moment $\textbf{m}$. It can be further assumed that the charges move in a plane with its normal vector tilted by an angle (defined by the helix radius and the pitch) with respect to the $z$-axis. Therefore, the induced magnetic dipole moment must, be tilted respectively. Furthermore, when the electric current is largest, the electric charge is separated predominantly along the direction along the $x$-direction. This temporal charge distribution at exactly $\omega t=\pi/2$ causes the ellipicity of $\textbf{p}$. Similarly, when $\omega t$ approaches $2\pi$, the decaying current at the ends of the nanohelix induces the small ellipticity to $\textbf{m}$. \\
It is clear now that both electric and magnetic dipoles have components in all directions. In particular, we note here that transverse components are necessary to excite the fundamental resonance with the incoming plane-wave-like (transverse) RCP field at normal incidence. For a fixed orientation of the nanohelix in space, the two dipoles do not change. Thus, as an intrinsically chiral three-dimensional structure has to preserve its handedness regardless of the point of observation, the resonance can be excited regardless of the direction of the incoming RCP \cite{fruhnert_2017}. Importantly, both dipoles are driven by the same source which is $\pi/2$ delayed for each dipole, namely the harmonically oscillating electric charge along a finite helical wire. At the wavelength of the fundamental resonance, the nanohelix can be equivalently described as a point-like scatterer driven by coupled electric and magnetic dipoles $\textbf{p}+i\textbf{m}$. In addition, the structure can also be described as a nanoLC-circuit accessible for only one handedness of the incoming light (here RCP) with regard to its own geometrical twist \cite{jaggard_1979}.\\
Although the geometry does not allow the scatterer to support electric and magnetic dipoles being entirely parallel to each other, the two moments share predominantly a common direction of oscillation (see Figs.~\ref{fig:_dipoles_rcp} and \ref{fig:_resonance_1_rcp}a). Therefore, the intrinsic cross polarizability $G''\sim \Im\left[\textbf{p}^{*}\cdot\textbf{m}\right]$ \cite{hu_2017} is non-zero. Note that the demanded non-zero projection of the electric and magnetic dipoles onto each other fulfills the condition of a true chiral entity according to Barron's definition \cite{barron_2009}. In a first approximation, the value of $G''$ can be estimated by $\text{p}_{z}+i\text{m}_{z}$ which is, hence, responsible for the optical activity of the nanohelix. A point-like system of coupled electric and magnetic dipoles oscillating along the same direction, which we call here a chiral dipole, is parity-asymmetric and time-symmetric (see Supplementary Information). The time delay of $\pi/2$ between the electric and magnetic dipoles ($\text{p}_{z}+i\text{m}_{z}$) resembles the relation between the electric and magnetic field in a chiral medium $\textbf{H}=\pm i\sqrt{\epsilon/\mu}\textbf{E}$, with $\epsilon$ and $\mu$ being the permittivity and permeability of the medium, respectively. Sets of other dipole components of significant strength which oscillate along perpendicular directions (e.g., ($\text{p}_{y}$,$\text{m}_{z}$) or ($\text{p}_{z}$,$\text{m}_{y}$)) do not contribute to the optical activity as they are parity-invariant. Such a dipolar system can be found in a split-ring resonator under normal illumination for which the structure stays optically inactive \cite{plum_2009}. Consequently, the presented helix is indeed an intrinsically chiral entity. \\
To complete the analysis of the chiroptical response of the nanohelix at its fundamental resonance, we investigate the far-field scattered light at $\lambda=1450$ nm. The emission of a dipole placed near a dielectric interface, regardless of its type and orientation, is predominantly directed into the optically denser medium. In the chosen scheme of an air-glass interface, the scattered light peaks at the angle of $\text{NA}=1.0$ in the glass substrate \cite{lukosz1977_2}. Hence, for experimental observation, we image the back focal plane of the microscope objective in transmission, where the scattered light is angularly separated from the transmitted beam (see Methods and Refs.~\cite{neugebauer_2014,wozniak_2015} for more details). Figure~\ref{fig:_scattering} shows the acquired distribution of the light emitted by the nanohelix excited with RCP light. The total intensity exhibits a ring-like distribution. Polarization analysis of the scattered light further reveals that it is predominantly radially polarized and, hence, dominated by the radiation of the electric dipole \cite{wozniak_2015}, which is slightly tilted with respect to the $z$-axis. The emission of the induced magnetic dipole moment, thus, contributes extremely weakly to the intensity of the far-field scattered light. This is caused not only by the lower power radiated by the magnetic dipole but also by its lower transmissivity through the air-glass interface in comparison to the electric dipole \cite{lukosz_1977}. Nonetheless, the magnetic dipole is of utmost importance for the chiral response of the nanohelix as discussed above. The experimental data is in very good agreement with the analytical calculations of the far-field emission of electric and magnetic dipole moments above a plane dielectric interface. To this end, the amplitudes and phases of the FDTD-retrieved dipoles at $\lambda=1450$ nm and RCP light are used. The good overlap confirms again that the contribution of higher-order multipoles can be neglected for the investigated resonance. \\
\section{First higher-order resonance}
The dipole decomposition presented in Fig.~\ref{fig:_dipoles_rcp} indicates the first higher-order resonance of the nanohelix at $\lambda = 840$ nm. Similar to the fundamental resonance, the exited mode responds more strongly to RCP illumination (see Supplementary Information). Figure~\ref{fig:_resonance_2_rcp} shows the corresponding distributions of the charge and current densities retrieved from the FDTD simulations at two different points in time delayed by $\pi/2$. Because at a certain point in time the positive and negative charges accumulate at both ends of the helix and its middle, the excited electric dipole oscillates predominantly along the $x$-axis. Such charge distribution will produce two curl-like currents flowing towards the two ends of the nanohelix. Accordingly, each half-loop will produce two magnetic dipole moments of the same transverse direction but pointing in opposite $z$-direction. While the anti-parallel longitudinal components will cancel each other, the transverse components will interfere constructively. Effectively, the running charges will produce a predominantly $x$-polarized magnetic dipole. The dipolar decomposition (see Fig.~\ref{fig:_dipoles_rcp}) allows for describing the first higher-order resonance as a point-like system of coupled electric and magnetic dipoles of ellipticities of 0.9998 and 0.9991, respectively. Due to the different charge and current distributions in comparison to the fundamental resonance (see Fig.~\ref{fig:_resonance_1_rcp}a and \ref{fig:_resonance_2_rcp}a), the cross polarizability of the resonance can be estimated by the contribution of $\text{p}_{x}-i\text{m}_{x}$ to the scattering of the nanohelix. The transverse $x$-polarized components can be treated, hence, as the chiral dipole of the resonance and have a better coupling overlap with the transverse RCP at normal incidence.     
\section{Conclusion}
In summary, we have studied the optical response of an intrinsically chiral plasmonic nanostructure exhibiting sub-wavelength dimensions. In contrast to the majority of works discussed in the literature so far, we study the chiroptical response of a single-loop nanohelix to understand its interaction with circularly polarized light on the single-particle level. The discussion focuses on the fundamental and first higher-order resonances of the nanohelix. We investigated the chiroptical response by measuring and numerically calculating its differential absorption of right- and left-handed circularly polarized light. From the numerical data, we retrieved the multipolar response and determined the structure of the dominating electric and magnetic dipole moments. We found that the chirpotical response of the nanohelix is fundamentally different for each of the two resonances, as the charge distributions define chiral dipoles oriented in the $z$- and $x$-directions for the fundamental and first-high order resonances, respectively. The non-zero projection of the two dipoles onto each other is parity-asymmetric and time-invariant and, hence, responsible for the chirality of the structure and determine its strength. The experimentally observed far-field emission at the fundamental resonance is in a very good agreement with the one found analytically. Our findings contribute to the understanding of chiral light-matter interactions based on the multipolar response of a point-like chiral entity.
\begin{figure*}[t]
\centering 
\includegraphics[width=1\textwidth]{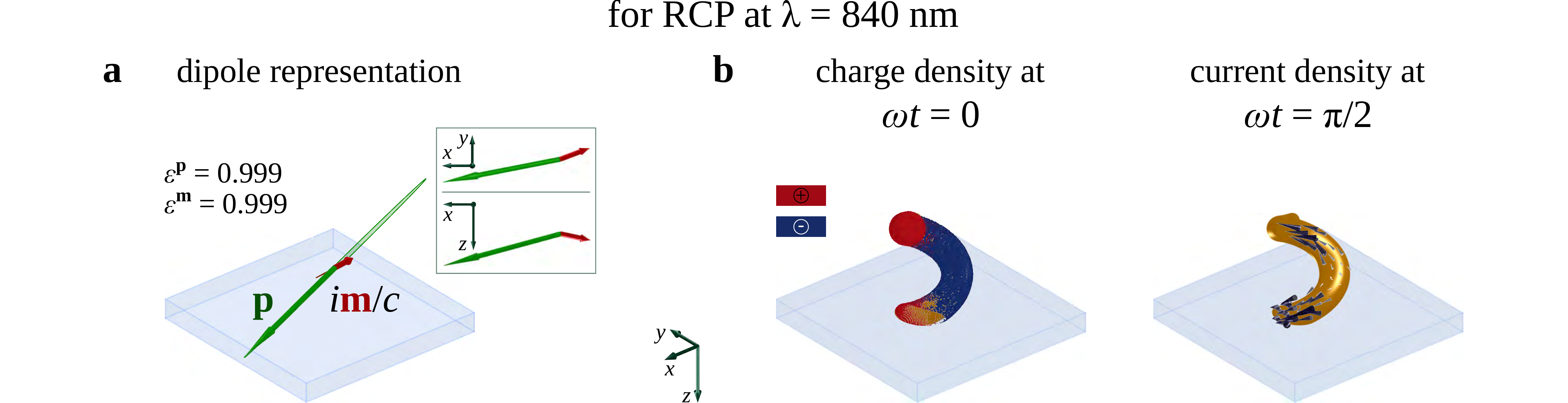}  
\caption{
\textbf{Dipolar representation of the first higher-order resonance.} Same as Fig.~\ref{fig:_resonance_1_rcp}, but for $\lambda = 840$ nm.}
\label{fig:_resonance_2_rcp}
\end{figure*}
\section{Methods}
\noindent \footnotesize{\textbf {Helix fabrication.} A helix of nanometric dimensions was fabricated using electron-beam-induced deposition (EBID) \cite{hoflich_2011}. For that purpose, a 170 $\upmu$m thick borosilicate glass cover slip with a thin layer (46 nm) of indium tin oxide (ITO) for sufficient surface conductivity was used as a substrate. The EBID process took place in a vacuum chamber of a dual-beam instrument FEI Strata DB 235. The metal-organic precursor dimethyl-gold(III)-acetylacetonate (Me$_{2}$Au(acac)) was heated up to 34$^\circ$C and introduced to the chamber by a gas-injection system, placing the inlet needle approximately 0.5 mm above the sample. In the proximity of a focused electron beam, the injected precursor molecules are dissociated and their non-volatile parts form a deposit while the volatile parts are pumped out. By changing the relative position of the focus with respect to the substrate, a single-loop helix was produced. For the chosen dimensions, a pixel spacing of 1 nm and a dwell time of 1.2 ms lead to vertical pitch heights of around 230 nm. Deposition process was carried out under pressure of 0.2 - 0.5 mPa at an acceleration voltage of 15 kV and a current below 100 pA, resulting in a helix wire with diameter of 20 nm. As EBID process does not allow to fabricate purely metallic structures \cite{wozniak_2015_2}, the helix was subsequently coated with gold using electron beam evaporation under a pressure of 0.1 mPa. Conformity of the coat was achieved by glancing angle deposition under $87^{\circ}$ tilt and rotating the sample. The resulting core-shell helix (similar to structures reported recently Kosters \textit{et al.} in Ref.~\cite{kosters_2017}) had a diameter of approximately 70 nm \cite{haverkamp_2017}. Lastly, the thin gold and ITO layers around the helix were removed by an Ga focused ion-beam operating at  a voltage of 30 kV and a beam current of 10 pA.\\

\noindent \textbf{Experimental setup and measurement scheme.} The custom-build measurement system (similar to the one used in Refs.\cite{banzer_2010,neugebauer_2014,wozniak_2015,banzer_2016}) was equipped with a broadband light source, which covers the spectral range of interest between 1300~nm and 1650~nm. To provide equal illumination condition at each wavelength, the measurements were performed wavelength-by-wavelength with a spectral step of 10 nm (effectively the experimentally measured spectra consist of data points for 36 wavelength). To this end, the laser Gaussian beam was filtered spectrally by an acousto-optical tunable filter down to approximately 5 nm spectral width. The quasi-monochromatic beam was further polarized with a set of a linear polarizer and an achromatic quarter-wave plate, and subsequently focused with a microscope objective with a high NA of 0.9. The incoming beam only partially filled the back focal plane of the microscope lens, reducing the effective focusing $\text{NA}$ down to 0.5 and preventing formation of strong longitudinal field components across the focal plane. The focus diameter changed slightly for individual wavelengths, leading to an insignificant increase of the energy density on the optical axis for shorter wavelengths. During the measurement the nanohelix is placed on the optical axis in the focal plane. The glass substrate with the investigated nanohelix on top was placed on a three-dimensional piezo-stage. This enables precise positioning of the sub-wavelength structure on the optical axis in the focal plane. The actual on-axis position is found by raster-scanning the nanohelix across the focused beam, and by a subsequent analysis of the scans. The reflected and the back-scattered light is collected by the focusing lens. The smaller waist of the incoming beam with regard to the diameter of the back focal plane aperture resulted in a plane-wave-like illumination of the structure at normal incidence. On the other hand, the back-scattered and reflected light could be collected within the full angular range provided by the numerical aperture (0.9) of the objective. For separation of the incident and the reflected light, a circular-polarization-preserving beam-splitter-system \cite{tidwell_1992} was inserted between the source and the focusing lens. The transmitted and the forward-scattered light was collected by a second immersion-type objective of higher $\text{NA 1.3}$. In both directions, the total power was measured with photo-diodes. The acquired data was normalized to the power of the impinging light onto the sample by measuring a reference signal for the air-glass interface and by calculating the corresponding power Fresnel coefficients for each wavelength for the chosen illumination scheme. For experimental observation of the forward-scattered light in the far-field, the photo-diode in transmission was replaced by a lens and a camera to image the back focal plane of the objective. Due to the very high NA of 1.3 of the microscope lens, the scattered light could be clearly distinguished from the transmitted beam within the angular range of $\text{NA} \in[0.9,1.3]$. For details see Refs.~\cite{neugebauer_2014,wozniak_2015}. \\

\noindent \textbf{Numerical modeling.} The theoretical transmittance and reflectance spectra were obtained using a Maxwell's-equations solver based on the finite-difference time-domain method (Lumerical FDTD Solutions). All simulations were carried out using a Gaussian beam with the same characteristics as those used in the experiments. The permittivity dispersion of gold was taken from Ref.~\cite{johnson_1972}. Based on ellipsometry measurements, the refractive index of the glass substrate was set to 1.51, and its dispersion was neglected. For equal illumination conditions, the circularly polarized Gaussian beam was set to be monochromatic. The simulations were run in the spectral range between 450 nm and 1900 nm with a spectral step size of 10 nm (see Supplementary Information). The focusing scheme was chosen in accordance with the experiment (same focusing NA of 0.5 for all wavelengths) and, hence, the size of the focal spot changed slightly as explained above. The simulation provided the densities of charge and current as well as the polarization density \textbf{P} for the subsequent spectral multipole decomposition of the investigated nanohelix.\\

\noindent \textbf{Multipolar decomposition.} The FDTD-retrieved spatial distributions of the polarization density $\textbf{P}(\textbf{r})$ in close proximity of the nanohelix was further evaluated to determine the spectral multipolar response using \cite{stratton_1941,evlyukhin_2016}:
\begin{equation*}
\begin{aligned} 
	\textbf{p} &= \int\textbf{P}(\textbf{r})\text{ }d\textbf{r}\text{,} \\
	\textbf{m} &= -\frac{i\omega}{2}\int\left[\textbf{r}\times\textbf{P}(\textbf{r})\right]\text{ }d\textbf{r}\text{,} \\
	\textbf{Q}^\text{e} &= 3\int\left[\textbf{r}\textbf{P}(\textbf{r})+\textbf{P}(\textbf{r})\textbf{r}-\frac{2}{3}\left[\textbf{r}\cdot\textbf{P}(\textbf{r})\right]\textbf{U}\right]d\textbf{r}\text{,} \\
	\textbf{Q}^\text{m} &= \frac{\omega}{3i}\int\left(\left[\textbf{r}\times\textbf{P}(\textbf{r})\right]\textbf{r}+\textbf{r}\left[\textbf{r}\times\textbf{P}(\textbf{r})\right]\right)d\textbf{r}\text{,}
	\label{eq:_scattering_coefficients}
\end{aligned}
\end{equation*}
 where $\textbf{p}$, $\textbf{m}$, $\textbf{Q}^{e}$ and $\textbf{Q}^{m}$ represent the electric and magnetic dipole and quadrupole moments, respectively, $\textbf{r}=(x\text{,}y\text{,}z)$ is a position in space and $\textbf{U}$ is the 3 x 3 unit tensor. It shall be noted that the equations above are valid for a scatterer of very small dimensions compared to the excitation wavelength. The helix presented in Fig.~\ref{fig:_schematics} fulfills the condition as the pitch (the largest dimension of the structure) is over six times smaller and almost four times smaller than the wavelength of the fundamental resonance and the first higher-order resonance, respectively. For scatterers of larger dimensions, the exact solution, which takes into account the so-called toroidal moments, has been recently presented by Alaee \textit{et al.} in Ref.~\cite{alaee_2018}. \\

\noindent \textbf{Far-field emission into substrate.} For a point (dipolar) emitter in air close to a dielectric interface, the far-field intensity distribution into the substrate reads $\text{I}=\left|\text{E}_\text{TM}\right|^{2}+\left|\text{E}_\text{TE}\right|^{2}$ with \cite{novotny_2012} 
\begin{equation*}
\begin{aligned} 
	\text{E}_\text{TM} &= \text{E}_\text{TM}^\textbf{p}+\text{E}_\text{TM}^\textbf{m}\text{,} \\[5pt]
	\text{E}_\text{TE} &= \text{E}_\text{TE}^\textbf{p}+\text{E}_\text{TE}^\textbf{m}{,}
\end{aligned}
\end{equation*}
 where $\text{E}_\text{TM}$ and $\text{E}_\text{TE}$ represent two polarizations of the scattered light for which the electric field oscillates parallelly and perpendicularly with respect to the meridional plane, respectively. The contribution of the electric dipole $\textbf{p}=\left(\text{p}_{x}\text{, p}_{y}\text{, p}_{z}\right)$  to the far-field mission can be determined from
\begin{equation*}
\begin{aligned} 
\left( \begin{array}{c} \text{E}_\text{TM}^\textbf{p} \\[5pt] \text{E}_\text{TE}^\textbf{p} \end{array} \right) &= C\textbf{F}_{t}\textbf{M}_\textbf{p}\left( \begin{array}{c} \text{p}_\textit{x} \\ \text{p}_\textit{y} \\ \text{p}_\textit{y} \end{array} \right)
\end{aligned}
\end{equation*}
and of the magnetic dipole $\textbf{m}=\left(\text{m}_{x}\text{, m}_{y}\text{, m}_{z}\right)$ reads
\begin{equation*}
\begin{aligned} 
\left( \begin{array}{c} \text{E}_\text{TM}^\textbf{m} \\[5pt] \text{E}_\text{TE}^\textbf{m} \end{array} \right) &= \frac{C}{c_{0}}\textbf{F}_{t}\textbf{M}_\textbf{m}\left( \begin{array}{c} \text{m}_\textit{x} \\ \text{m}_\textit{y} \\ \text{m}_\textit{y} \end{array} \right)
\end{aligned}
\end{equation*}
with 
\begin{equation*}
\begin{aligned} 
	C &= e^{i\left|\textbf{k}\right|nr}\frac{\left|\textbf{k}\right|^{2}\sqrt{\left|\textbf{k}\right|^{2}n^{2}-\text{k}_\bot^{2}}}{4\pi r\epsilon_{0}\kappa_{z}}e^{i\kappa_{z}d}\text{.} \\[5pt]
\end{aligned}
\end{equation*}
Here, $\textbf{k}=\left(\text{k}_{x}\text{, k}_{y}\text{, k}_{z}\right)$ represents the wave vector in free space ($\left|\textbf{k}\right|=2\pi/ \lambda$), $\text{k}_\bot=\left(\text{k}_{x}^{2}+\text{k}_{y}^{2}\right)^{1/2}$ is the transverse wave number in free space, $\kappa_{z}=\left(\left|\textbf{k}\right|^{2}-\text{k}_\bot^{2}\right)^{1/2}$ and $n$, $\epsilon_{0}$, $c_{0}$, $d$ and $r$ are the refractive index of the glass substrate, the vacuum permittivity, the speed of light, the distance between the point emitter and the substrate, and the distance to the observer, respectively. In the equations above $\textbf{F}_{t}$ stands in for the transmission matrix defined as
\begin{equation*}
\begin{aligned} 
\textbf{F}_{t} &= \left( \begin{array}{cc} t_\text{TM} & 0 \\ 0 & t_\text{TE} \end{array} \right)
\end{aligned}
\end{equation*}
 with  $t_\textit{TM}$ and  $t_\textit{TE}$ being the Fresnel transmission coefficients for TM- and TE-polarized light. The two matrices $\textbf{M}_\textbf{p}$ and $\textbf{M}_\textbf{m}$ are the rotation matrices which project the electric field of the light emitted by the dipoles onto the plane-waves of the angular spectrum of the back focal plane and are defined as
\begin{equation*}
\begin{aligned} 
\textbf{M}_\textbf{p} &= \left( \begin{array}{ccc} \frac{\text{k}_{x}\kappa_{z}}{\text{k}_\bot \left|\textbf{k}\right|} & \frac{\text{k}_{y}\kappa_{z}}{\text{k}_\bot \left|\textbf{k}\right|}  & -\frac{\text{k}_\bot}{\left|\textbf{k}\right|} \\[5pt]
-\frac{\text{k}_{y}}{\text{k}_\bot} & \frac{\text{k}_{x}}{\text{k}_\bot} & 0 \end{array} \right) \\[5pt]
\textbf{M}_\textbf{m} &= \left( \begin{array}{ccc} -\frac{\text{k}_{y}}{\text{k}_\bot} & \frac{\text{k}_{x}}{\text{k}_\bot} & 0  \\[5pt]
 -\frac{\text{k}_{x}\kappa_{z}}{\text{k}_\bot \left|\textbf{k}\right|} & -\frac{\text{k}_{y}\kappa_{z}}{\text{k}_\bot \left|\textbf{k}\right|}  & \frac{\text{k}_\bot}{\left|\textbf{k}\right|}\end{array} \right)\text{.}
\end{aligned}
\end{equation*}
Accordingly, the $x$- and $y$-polarized components of the electric field in the far-field read $\text{E}_{x}=\text{E}_\text{TM}\cos\phi-\text{E}_\text{TE}\sin\phi$ and $\text{E}_{y}=\text{E}_\text{TM}\sin\phi+\text{E}_\text{TE}\cos\phi$, where the angle $\phi$ is the azimuth coordinate measured from the $x$-axis towards $y$-axis of the coordinate system whose center coincides with the point emitter (see Fig.~\ref{fig:_resonance_1_rcp}). } 
%
\section{Acknowledgments}
PW acknowledges fruitful discussions with Dr. S. Nechayev. IDL and PB acknowledge the support from CONACyT -- DAAD (Proalmex) grant under the project No. 267735. PB acknowledges support by the Alexander von Humboldt Foundation and the Federal Ministry of Education and Research. CH and KH acknowledge funding from the Helmholtz Association within the Helmholtz Postdoc Program.
\bibliography{biblioteca}

\newpage
\onecolumngrid

\begin{tcolorbox}
\centerline{\Large{Chiroptical response of a single plasmonic nanohelix}}
\vspace{5 mm}
\centerline{\large{SUPPLEMENTARY INFORMATION}}
\end{tcolorbox}

\renewcommand{\thefigure}{S\arabic{figure}}
\setcounter{figure}{0}

\section{Spectral multipole expansion of an individual single-loop plasmonic helix}
\begin{figure}[t]
	\centering 
	\includegraphics[width=1\textwidth]{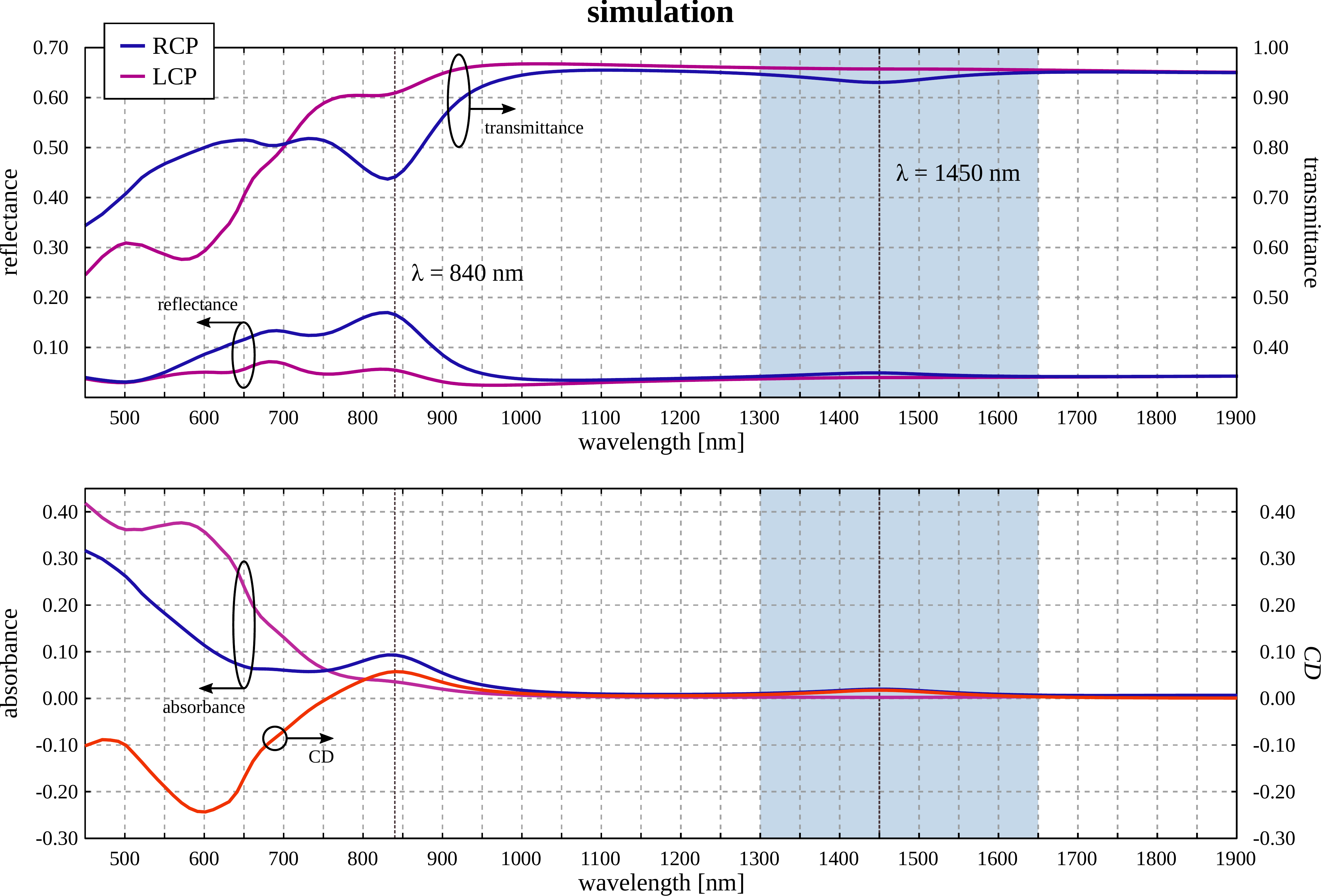} 
	\caption{Broadband FDTD simulations of the chiroptical response of the investigated nanohelix. Dispersion of the differential absorption ($CD$) for right- (RCP) and left-circularly polarized light (LCP) by the helix depicted in Fig.~1a in the main text. The absorbance of light is determined from the acquired reflectance and transmittance spectra in the spectral range between 450 nm and 1900 nm. The first two $CD$ resonances are determined to be at 1450 nm and at 840 nm. The blue-shaded spectral range spans over the spectral range of the experiment.} 
	\label{fig:_si_spectra}
\end{figure}

\begin{figure}[t]
	\centering 
	\includegraphics[width=1\textwidth]{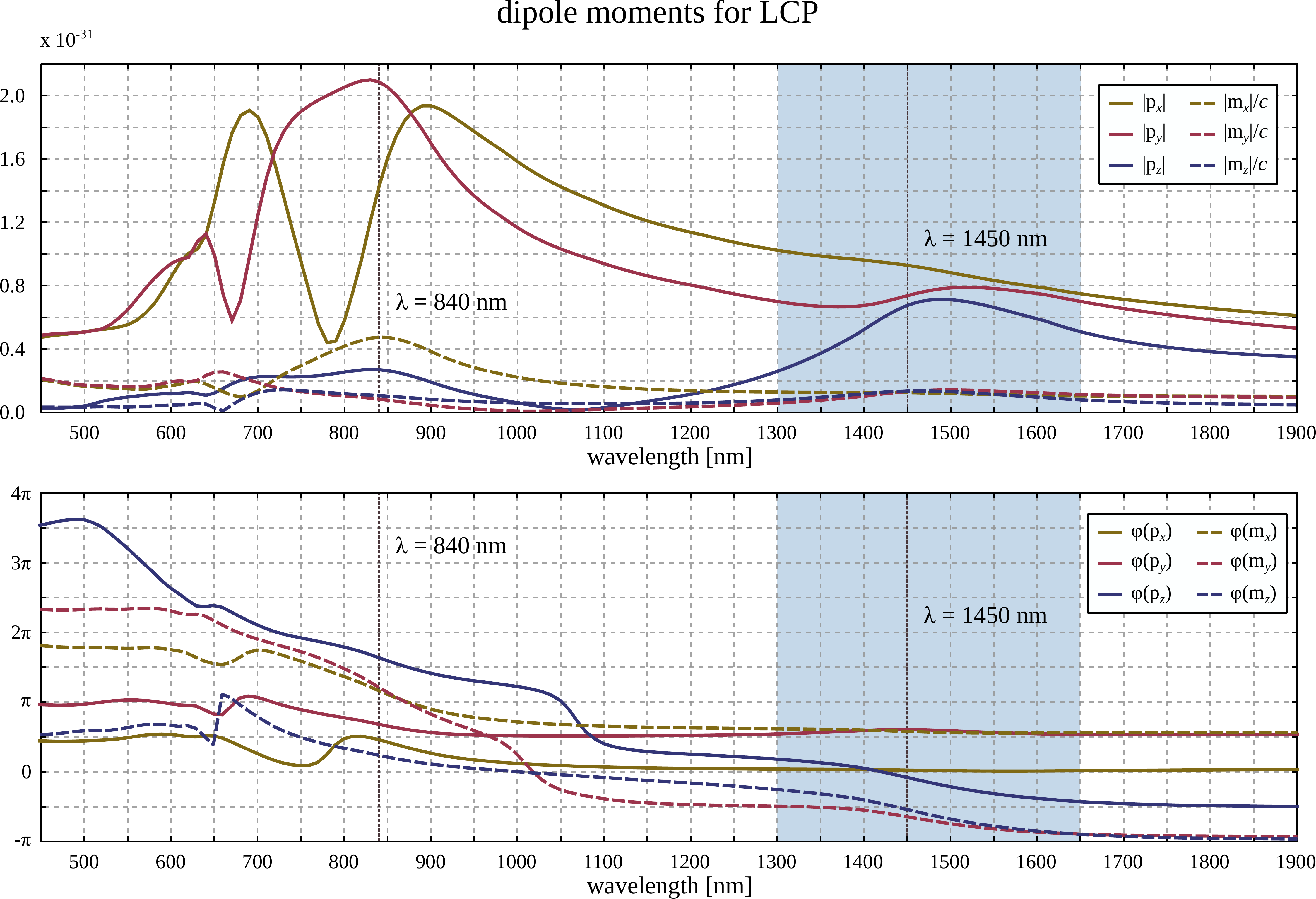} 
	\caption{Absolute values and phases of the components of the electric $\textbf{p}$ and magnetic $\textbf{m}$ dipole moments for the right-handed helix under LCP illumination.} 
	\label{fig:_si_dipoles_lcp}
\end{figure}

\begin{figure}[t]
	\centering 
	\includegraphics[width=1\textwidth]{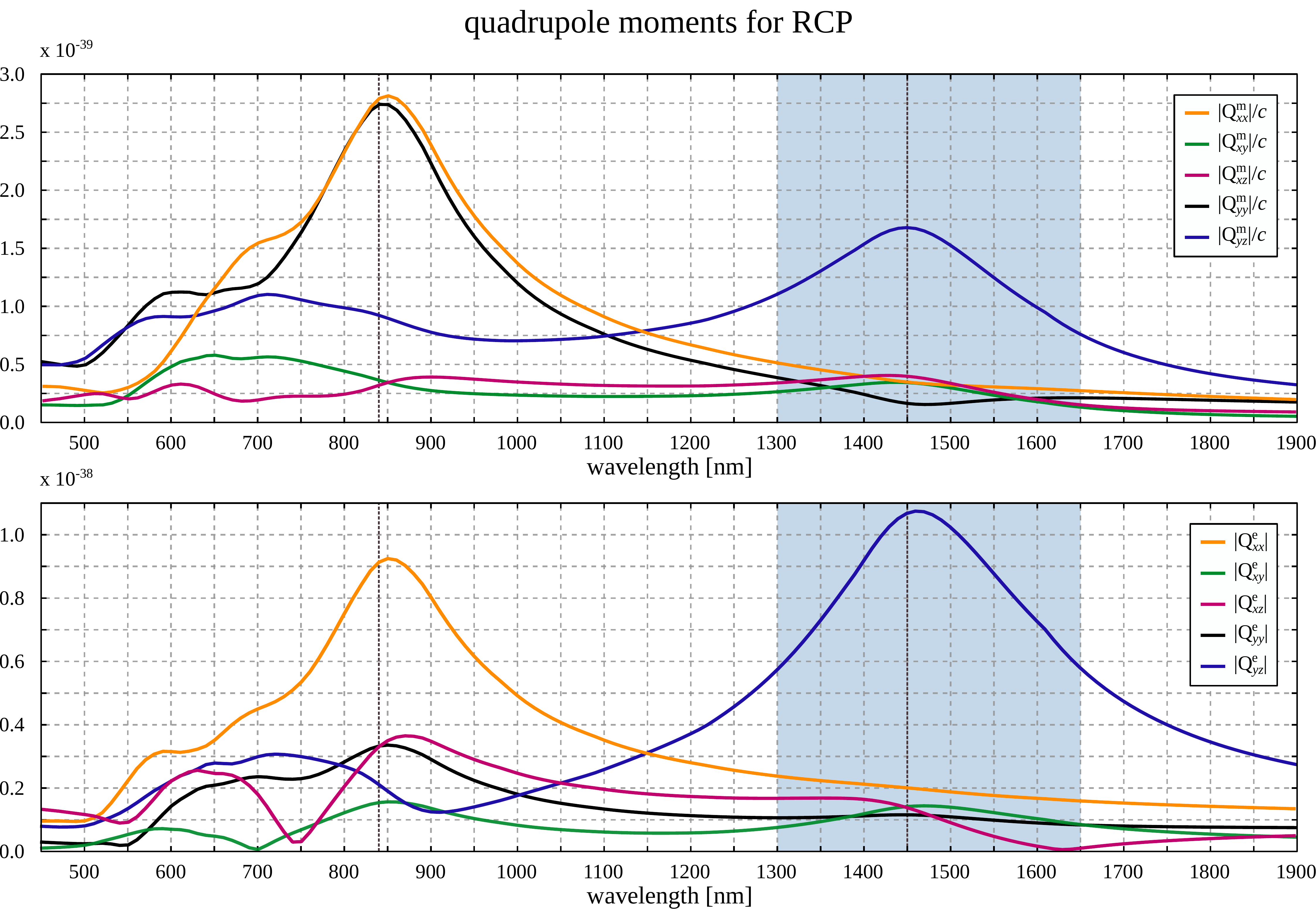} 
	\caption{Absolute values of the components of the magnetic $\textbf{Q}^\text{m}$ and electric $\textbf{Q}^\text{e}$ quadrupoles of the helix under RCP illumination.} 
	\label{fig:_si_quadrupoles_rcp}
\end{figure}

\begin{figure}[t]
	\centering 
	\includegraphics[width=1\textwidth]{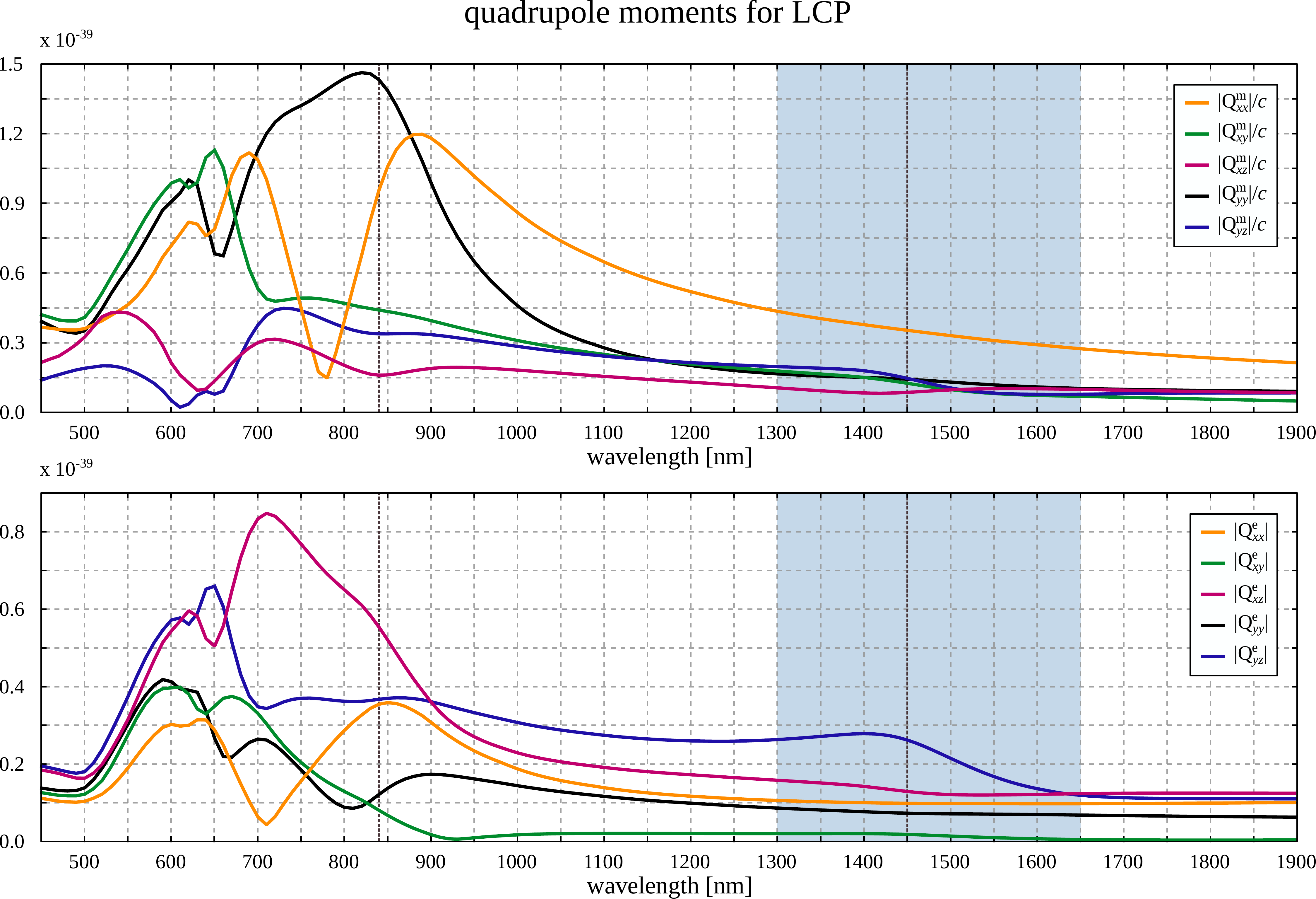} 
	\caption{The same as Fig.~\ref{fig:_si_quadrupoles_rcp}, but for an LCP excitation.} 
	\label{fig:_si_quadrupoles_lcp}
\end{figure}

The FDTD simulations were performed for a spectral range between 450 nm and 1900 nm with a step size of 10 nm. The nanohelix was excited with a weakly focused ($\text{NA}=0.5$) circularly polarized Gaussian beam following the illumination scheme in the experiment. Fig.~\ref{fig:_si_spectra} presents the reflectance, transmittance, absorbance, and differential absorption spectra for RCP and LCP light. The reflectance and the transmittance spectra are normalized to the input power and the two other quantities are calculated as explained in the main text. Accordingly, the experimentally investigated resonance at 1450 nm is identified to correspond to the fundamental mode of the structure and the first higher-order resonance is expected at the wavelength of 840 nm. Furthermore, using the FDTD-recorded polarization density $\textbf{P}$, the spectral multipolar decomposition was done as described in Methods. The optical response of the helix was expanded into electric and magnetic multipoles up to order 3 (dipole, quadrupole, and octupole). The investigated structure can be effectively described with dispersive electric and magnetic dipoles as depicted in Figs.~\ref{fig:_si_dipoles_lcp}-\ref{fig:_si_quadrupoles_lcp}. Due to the very weak strength of the octupoles ($\sim$11-12 orders of magnitude weaker than the retrieved quadrupoles), only the spectra for the dipoles and quadrupoles are presented.
\section{\textit{PT}-symmetries of\\a system of coupled electric and magnetic dipoles}
\begin{figure}[t]
	\centering 
	\includegraphics[width=1\textwidth]{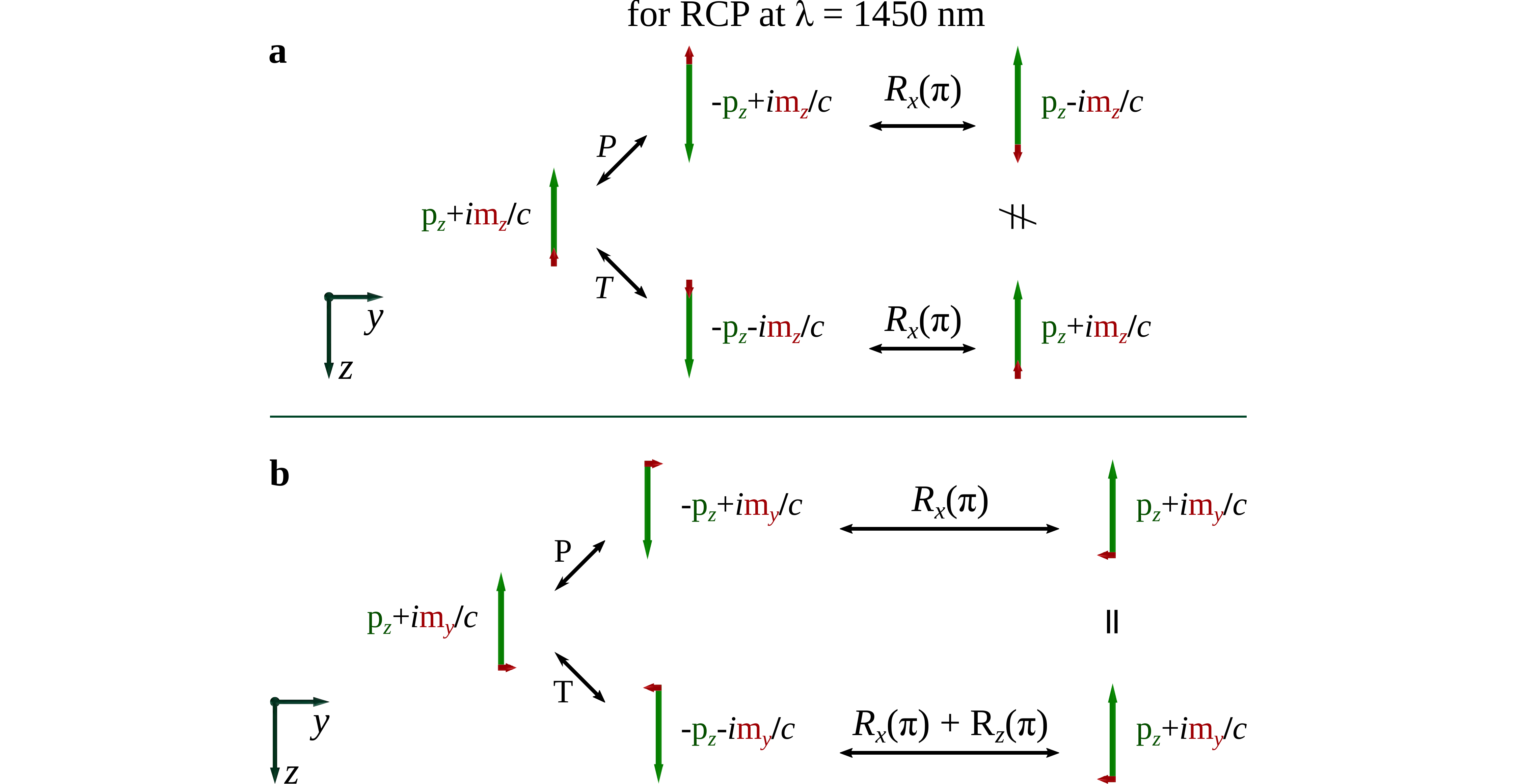} 
	\caption{Parity ($P$) and time ($T$)  inversions of a point-like system of coupled electric and magnetic dipoles of (\textbf{a}) parallel and (\textbf{a}) perpendicular orientations.}
	\label{fig:_si_pt_symmetry}
\end{figure}

Figure~\ref{fig:_si_pt_symmetry}a presents the chiral dipole of the fundamental resonance $\text{p}_{z}+i\text{m}_{z}$ under a parity transformation ($P$) and time inversion ($T$). This can be seen as a test for chirality of the $z$-polarized coupled electric and magnetic dipoles \cite{barron_2009}. While an electric dipole has an odd parity, a magnetic dipole, in contrast, exhibits even parity. A \textit{P} transformation hence results in a dipole system defined by $-\text{p}_{z}+i\text{m}_{z}$ which does not coincide with the original set of dipoles under any isometric transformation except $P^{-1}$. Effectively, the $P$ transformed dipole system describes a left-handed structure, which does not coincide with the investigated right-handed nanohelix. Moreover, a true three-dimensional chiral system is time (\textit{T}) irreversible. The sense of rotation of the eddy current (at $\omega t=\pi/2$) changes under a \textit{T} transformation ($\omega t\rightarrow-\omega t$) and the magnetic dipole moment alters its orientation respectively. If $\text{p}_{z}$ and $\text{m}_{z}$ were not coupled via the geometry of the nanohelix, the longitudinal electric dipole moment would stay unaffected. Nevertheless, the time-harmonic oscillation of charge forces $\text{p}_{z}$ to reorient with $\text{m}_{z}$, and the resulting state is again indistinguishable from the initial one despite a $\pi$ phase change ($\textit{P}(\text{p}_{z}+i\text{m}_{z})\neq\text{p}_{z}+i\text{m}_{z}=\textit{T}(\text{p}_{z}+i\text{m}_{z})$). The simultaneous change of both dipole moments under \textit{T} inversion is a consequence of the magnetoelectric coupling induced by the geometry of the nanostructure. \\
On the contrary, a point-like system of coupled electric and magnetic dipoles oscillating along two orthogonal directions does not fulfill the argument of the $PT$-symmetries. Figure~\ref{fig:_si_pt_symmetry}b shows another set of dipole components $\text{p}_{z}+i\text{m}_{x}$ at $\lambda=1450$ nm presented in Fig. 3 in the main text. Under $P$ transformation the resulting system can be easily superimposed with the initial pair of dipoles by a rotation about the $x$-azis ($R_{x}(\pi)$). Also, rotations about the $x$- and $z$-axises ($R_{x}(\pi)+R_{z}(\pi)$) of the $T$-transformed dipoles can be overlapped with the input set of dipoles. Since $\textit{P}(\text{p}_{z}+i\text{m}_{x})=\textit{T}(\text{p}_{z}+i\text{m}_{x})=\text{p}_{z}+i\text{m}_{x}$, a point-like system based on coupled orthogonal electric and magnetic dipoles is not chiral. \\
A point-like system of coupled electric and magnetic orthogonal dipoles is supported by a planar and solid split-ring resonator (SRR). Excited at normal incidence with, e.g., linearly or azimuthally polarized beam \cite{katsarakis_2004,banzer_2010,petschulat_2010,kuznetsov_2014}, the structure can be represented similarly as depicted in Fig.~\ref{fig:_si_pt_symmetry}b and, hence, a planar SRR does not exhibit optical activity. The excited dipoles can, however, be forced to have a common direction of oscillation, if the structure will be excited at oblique incidence. For tilted illumination, the wave vector of the incoming light and the SRR will create a chiral triad \cite{plum_2009}. Effectively the structure will appear to be a three-dimensional figure and exhibit so-called extrinsic chirality; hence extrinsic $G''\neq0$. 

\end{document}